\documentclass[12pt,epsf]{article}
\usepackage{graphicx}
\begin{document}

\renewcommand{\thefootnote}{\#\arabic{footnote}}
\setcounter{footnote}{0}


\def\a{\alpha}
\def\b{\beta}
\def\c{\varepsilon}
\def\d{\delta}
\def\e{\epsilon}
\def\f{\phi}
\def\g{\gamma}
\def\h{\theta}
\def\k{\kappa}
\def\l{\lambda}
\def\m{\mu}
\def\n{\nu}
\def\p{\psi}
\def\q{\partial}
\def\r{\rho}
\def\s{\sigma}
\def\t{\tau}
\def\u{\upsilon}
\def\v{\B}
\def\w{\omega}
\def\x{\xi}
\def\y{\eta}
\def\z{\zeta}
\def\D{\Delta}
\def\G{\Gamma}
\def\H{\Theta}
\def\L{\Lambda}
\def\F{\Phi}
\def\P{\Psi}
\def\S{\Sigma}

\def\o{\over}
\newcommand{\gsim}{ \mathop{}_{\textstyle \sim}^{\textstyle >} }
\newcommand{\lsim}{ \mathop{}_{\textstyle \sim}^{\textstyle <} }
\newcommand{\vev}[1]{ \left\langle {#1} \right\rangle }
\newcommand{\bra}[1]{ \langle {#1} | }
\newcommand{\ket}[1]{ | {#1} \rangle }
\newcommand{\EV}{ {\rm eV} }
\newcommand{\KEV}{ {\rm keV} }
\newcommand{\MEV}{ {\rm MeV} }
\newcommand{\GEV}{ {\rm GeV} }
\newcommand{\TEV}{ {\rm TeV} }
\def\diag{\mathop{\rm diag}\nolimits}
\def\Spin{\mathop{\rm Spin}}
\def\SO{\mathop{\rm SO}}
\def\O{\mathop{\rm O}}
\def\SU{\mathop{\rm SU}}
\def\U{\mathop{\rm U}}
\def\Sp{\mathop{\rm Sp}}
\def\SL{\mathop{\rm SL}}
\def\tr{\mathop{\rm tr}}

\def\IJMP{Int.~J.~Mod.~Phys. }
\def\MPL{Mod.~Phys.~Lett. }
\def\NP{Nucl.~Phys. }
\def\PL{Phys.~Lett. }
\def\PR{Phys.~Rev. }
\def\PRL{Phys.~Rev.~Lett. }
\def\PTP{Prog.~Theor.~Phys. }
\def\ZP{Z.~Phys. }
\newcommand{\bear}{\begin{array}}  \newcommand{\eear}{\end{array}}
\newcommand{\bea}{\begin{eqnarray}}  \newcommand{\eea}{\end{eqnarray}}
\newcommand{\beq}{\begin{equation}}  \newcommand{\eeq}{\end{equation}}
\newcommand{\bef}{\begin{figure}}  \newcommand{\eef}{\end{figure}}
\newcommand{\bec}{\begin{center}}  \newcommand{\eec}{\end{center}}
\newcommand{\non}{\nonumber}  \newcommand{\eqn}[1]{\beq {#1}\eeq}
\newcommand{\lmk}{\left(}  \newcommand{\rmk}{\right)}
\newcommand{\lkk}{\left[}  \newcommand{\rkk}{\right]}
\newcommand{\lhk}{\left \{ }  \newcommand{\rhk}{\right \} }
\newcommand{\del}{\partial}  \newcommand{\abs}[1]{\vert{#1}\vert}
\newcommand{\vect}[1]{\mbox{\boldmath${#1}$}}
\newcommand{\bib}{\bibitem} \newcommand{\new}{\newblock}
\newcommand{\la}{\left\langle} \newcommand{\ra}{\right\rangle}
\newcommand{\bfx}{{\bf x}} \newcommand{\bfk}{{\bf k}}
\newcommand{\gtilde} {~ \raisebox{-1ex}{$\stackrel{\textstyle >}{\sim}$} ~} 
\newcommand{\ltilde} {~ \raisebox{-1ex}{$\stackrel{\textstyle <}{\sim}$} ~}
\newcommand{\gtrsim}{ \mathop{}_{\textstyle \sim}^{\textstyle >} }
\newcommand{\lesssim}{ \mathop{}_{\textstyle \sim}^{\textstyle <} }
\newcommand{\ds}{\displaystyle}
\newcommand{\bi}{\bibitem}
\newcommand{\lar}{\leftarrow}
\newcommand{\rar}{\rightarrow}
\newcommand{\lrar}{\leftrightarrow}
\def\Frac#1#2{{\displaystyle\frac{#1}{#2}}}
\def\labelenumi{(\roman{enumi})}
\def\SEC#1{Sec.~\ref{#1}}
\def\FIG#1{Fig.~\ref{#1}}
\def\EQ#1{Eq.~(\ref{#1})}
\def\EQS#1{Eqs.~(\ref{#1})}
\def\lrf#1#2{ \left(\frac{#1}{#2}\right)}
\def\lrfp#1#2#3{ \left(\frac{#1}{#2}\right)^{#3}}

\newcommand{\fa}{F_{\cal A}}
\newcommand{\A}{{\cal A}}
\newcommand{\B}{{\cal B}}

\baselineskip 0.7cm

\begin{titlepage}

\begin{flushright}
\hfill DESY 06-034\\
\hfill hep-ph/0603265\\
\hfill April, 2006\\
\end{flushright}

\vskip 1.35cm
\begin{center}
{\large \bf
Gravitino Overproduction in Inflaton Decay
}
\vskip 1.2cm
Masahiro Kawasaki$^{1}$, Fuminobu Takahashi$^{1,2}$ and T. T. Yanagida$^{3,4}$
\vskip 0.4cm

${}^1${\it Institute for Cosmic Ray Research,
     University of Tokyo, \\Chiba 277-8582, Japan}\\
${}^2${\it Deutsches Elektronen Synchrotron DESY, Notkestrasse 85,\\
22607 Hamburg, Germany}\\
${}^3${\it Department of Physics, University of Tokyo,\\
     Tokyo 113-0033, Japan}\\
 ${}^4${\it Research Center for the Early Universe, University of Tokyo,\\
     Tokyo 113-0033, Japan}

\vskip 1.5cm

\abstract{ Most of the inflation models end up with non-vanishing
vacuum expectation values of the inflaton fields $\phi$ in the true
vacuum, which induce, in general, nonvanishing auxiliary field $G_\phi$ 
for the inflaton potential in supergravity. We show that
the presence of nonzero $G_\phi $ gives rise to  inflaton
decay into a pair of the gravitinos and are thereby severely constrained by 
cosmology especially if the gravitino is unstable and its mass is in a range of 
$O(100)$ GeV $\sim O(10)$ TeV.  For several inflation models, we explicitly
calculate the values of $G_\phi$ and find that most of them
are excluded or on the verge of being excluded for the gravitino mass in that range. 
We conclude that an inflation model with vanishing $G_\phi$, typically realized in
a chaotic inflation, is favored in a sense that it naturally avoids
the potential gravitino overproduction problem.  }
\end{center}
\end{titlepage}

\setcounter{page}{2}

\section{Introduction}
The recent Wilkinson Microwave Anisotropy Probe (WMAP) three year
data~\cite{Spergel:2006hy} is consistent with generic predictions in
inflation theories and hence strongly supports the basic idea of
inflationary universe.  Now we are reaching the stage to select the
specific inflation model that is favored by the observation.

In most of inflation models in supergravity (SUGRA) an inflaton field
$\phi$ has a non-vanishing expectation value at the potential
minimum~\cite{SUGRA-Inflation}.  This implies that the K\"{a}hler
potential for the inflaton field $\phi$ contains linear terms of the
inflaton field in the true vacuum, even if the minimal K\"{a}hler
potential is assumed at the beginning. That is, the inflaton
K\"{a}hler potential is written at the potential minimum as
$K=c\,\phi+c^*\phi^{\dagger}+ \phi\phi^{\dagger} + \cdots $.  With the
linear terms the inflaton field $\phi$ generically has a nonvanishing
auxiliary field $G_\phi$, where $G=K+{\rm ln}\,|W|^2$ (here and in what
follows a subscript $i$ denotes a derivative with respect to the field
$i$).  Here, $K$ and $W$ are the K\"{a}hler potential and
superpotential, respectively.  Recently, Ref.~\cite{Endo:2006zj} has
pointed out that, in the context of the moduli
problem~\cite{ModuliProblem}, the nonvanishing auxiliary field enables
the decay into a pair of the gravitinos to proceed with a rate much
higher than previously thought~\cite{Hashimoto:1998mu}.  We find this
decay process is (more) important in the reheating process of the
inflaton, and that a stringent constraint on the $G_\phi$ of the
inflaton potential must be satisfied to avoid an overproduction of the
gravitino keeping the success of the standard cosmology, especially when
the gravitino is unstable and its mass is in the range of 
$O(100){\rm\,GeV} - O(10){\rm\,TeV}$.  For several
inflation models, we explicitly calculate the values of $G_\phi$ to
exemplify how severe the bound is. Among the known inflation models, a
class of chaotic inflation models naturally avoids the potential gravitino
overproduction problem, since $G_\phi$ vanishes in the vacuum.

The new constraint on the inflaton potential $G_\phi$ depends on the
gravitino mass. In this letter, we restrict our discussion to the case
of the gravitino mass $O(1)$ TeV to make our point clear. We see how
severe the new constraint is compared with the cosmological constraint
on thermally produced gravitinos~\cite{Kawasaki:2004yh}.  Discussion
on more general cases of the gravitino mass is straightforward, which
will be given in~\cite{KTY}.

\section{Inflaton decay into a pair of gravitinos}
We estimate the decay rate of an inflaton field $\phi$ into a pair of
gravitinos.  To be concrete, we adopt the gravity-mediated SUSY
breaking scenarios, in which the gravitino mass is almost constant
during and after inflation.  The decay process we consider is
identical to that recently calculated for the modulus
decay~\cite{Endo:2006zj}.  The relevant interactions
are~\cite{WessBagger}
\bea
  e^{-1}\mathcal{L} &=&
  - \frac{1}{8} \epsilon^{\mu\nu\rho\sigma} 
  \left( G_\phi \partial_\rho \phi
    - G_{\bar \phi} \partial_\rho  \phi^\dag \right)
  \bar \psi_\mu \gamma_\nu \psi_\sigma
  \non\\&&
  - \frac{1}{8} e^{G/2} \left( G_{\phi} \phi + G_{\bar \phi}  \phi^\dag \right)
  \bar\psi_\mu \left[\gamma^\mu,\gamma^\nu\right] \psi_\nu, 
\label{eq:int}  
\eea
where $\psi_\mu$ is the gravitino field, and we have chosen the
unitary gauge in the Einstein frame with the Planck units, $M_P =1$.
The real and imaginary components of the inflaton field have the same
decay rate at the leading order~\cite{Endo:2006zj}:
\beq
\label{eq:inf2gravitino}
 \Gamma_{3/2} \equiv  \Gamma(\phi \rightarrow 2\psi_{3/2}) \simeq
  \frac{1}{288\pi} \frac{|G_\phi|^2}{g_{\phi \bar \phi}} \frac{m_\phi^5}{m_{3/2}^2 M_P^2}, 
\eeq
where we have assumed that the inflaton mass is much larger than the
gravitino mass: $m_\phi \gg m_{3/2}$, and $g_{ij^*}=K_{ij^*}$ is the
K\"ahler metric.  Note that $g_{\phi \bar\phi}=1$ for a canonically
normalized inflaton field.  Thus the decay rate is enhanced by the
gravitino mass in the denominator, which comes from the longitudinal
component of $\psi$ as emphasized in Ref.~\cite{Endo:2006zj}.

It should be noted that the above expression for the decay rate cannot
be applicable for $H>m_{3/2}$~\cite{ET}.  The decay proceeds only if
the Hubble parameter $H$ is smaller than the gravitino mass, since the
chirality flip of the gravitino forbids the decay to proceed
otherwise.  Intuitively, the gravitino is effectively massless as long
as $H>m_{3/2}$.

We should clarify another important issue: what is the longitudinal
component of the gravitino (i.e. goldstino) made of ? Similar issue
was discussed in the context of the non-thermal `gravitino' production
during preheating~\cite{Kallosh:1999jj}, and it was concluded that the
inflatino, instead of the gravitino in the low energy, was actually
created~\cite{Nilles:2001ry}~\footnote{
It should be noted, however, that the inflatinos produced during preheating
may be partially converted to the gravitinos in the low energy, since $G_\phi$ is generically
nonzero in the true minimum as we will show later.
This effect can further constrain the inflation models, but the detailed discussion is
beyond the scope of this letter.
}. The reason is that the `gravitino'
production occurs in a rather early stage of the reheating just
after the inflation ends, during which the energy stored in the
inflationary sector significantly contributes to the total SUSY
breaking.  In our case, however, the situation is completely
different; the decay into the gravitinos becomes effective, since we
consider a cosmological epoch, $H<m_{3/2}$, when the SUSY breaking
contribution of the inflaton is subdominant. Thus the gravitinos
produced by the inflaton through the above decay process should
coincide with those in the low energy. The gravitinos produced by
inflaton decay are genuine, and thereby the gravitino overproduction
problem is present.

\section{Cosmological constraint on $G_{\phi}$}
We restrict ourselves to the case of unstable gravitino with mass of $O(1)$
TeV. This mass region of the gravitino is an interesting region for
gravity-mediated SUSY breaking scenarios. In this case we have already a
cosmological constraint on the reheating temperature $T_R$ to avoid the
overproduction of gravitinos through thermal scattering~\cite{Kawasaki:2004yh}, that is~\footnote{
The expansion rate at the decay time is much smaller than the
gravitino mass $O(1)$ TeV and hence the condition $H<m_{3/2}$ is satisfied.
},
\beq
T_R < O(10^{6} \sim10^8) {\rm\, GeV},
\label{eq:trh-upper}
\eeq
depending on the gravitino mass and the hadronic branching ratio 
$B_h$.
This may be easily achieved with sufficiently small
couplings relevant for the inflaton decay. We assume that this bound on $T_R$ is
satisfied, which means the inflaton decay rate into the standard-model
particles must satisfy~\footnote{
Throughout this letter we require $\Gamma_{3/2} \ll \Gamma_{SM}$, otherwise the
standard cosmology would be upset.
}
\beq
\Gamma_{SM} \simeq \lrfp{\pi^2 g_*}{90}{\frac{1}{2}} \frac{T_R^2}{M_P} <  O(10^{-6} \sim 10^{-2}) {\,\rm GeV},
\label{eq:Trh-from-gp}
\eeq
where $g_* \sim 200$ counts the relativistic degrees of freedom. We point out that
too much gravitinos may be produced by the inflaton decay via the
interaction (\ref{eq:int}), even if the above
inequality is satisfied.  

The gravitino-to-entropy ratio is given by
\beq
Y_{3/2} \simeq 2 \,\frac{\Gamma_{3/2}}{\Gamma_{SM}}\frac{3}{4} \frac{T_R}{m_\phi},
\label{eq:ngs}
\eeq
where we have neglected the gravitino production from thermal scattering.
To keep the success of big-bang nucleosynthesis (BBN), the gravitino abundance 
must satisfy \cite{Kawasaki:2004yh}
\beq
m_{3/2} Y_{3/2} < O(10^{-14} \sim 10^{-11}) {\rm\,GeV}.
\label{eq:bbn}
\eeq
From (\ref{eq:Trh-from-gp}), (\ref{eq:ngs})  and (\ref{eq:bbn}), we obtain
\beq
\Gamma_{3/2} < O(10^{-23} \sim 10^{-17})
 \lrfp{m_{3/2}}{1{\rm\,TeV}}{-1}  \lrf{m_\phi}{10^6{\rm \,GeV}}{\rm\,GeV},
\eeq
or equivalently,
\beq
\label{eq:g-const}
\left|G_\phi \right| < O(10^{-4} \sim 10^{-1}) \lrfp{m_{3/2}}{1{\rm\,TeV}}{\frac{1}{2}}
   \lrfp{m_\phi}{10^6{\rm \,GeV}}{-2}.
\eeq

\begin{figure}[t]
\begin{center}
\includegraphics[width=10cm]{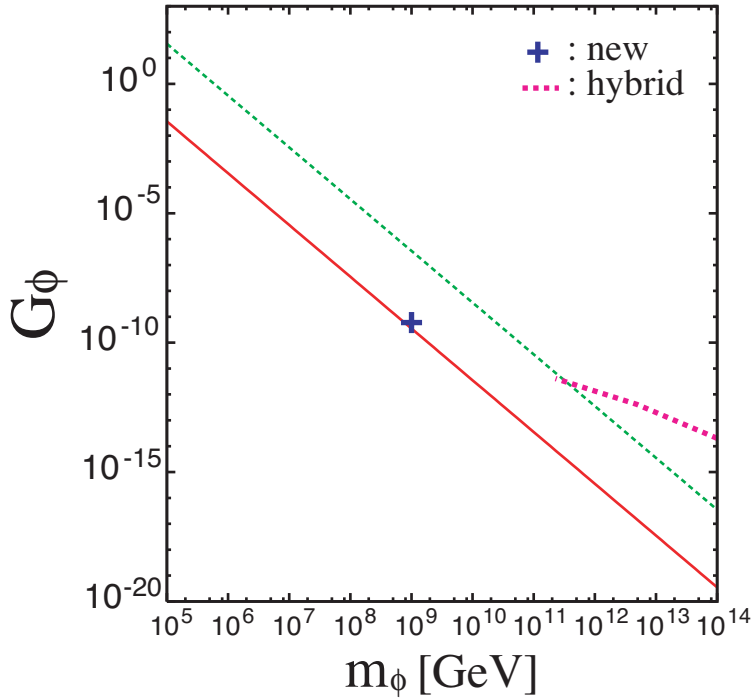}
\caption{Upper bound on the auxiliary field of the inflaton $G_\phi$
as a function of the inflaton mass $m_\phi$. The solid and dotted lines are for the hadronic
branching ratio $B_h = 1$ and $10^{-3}$, respectively. We set $m_{3/2} = 1{\rm\, TeV}$. 
The typical values of $G_\phi$ and $m_\phi$ for the new and hybrid inflation models
discussed in the text are also shown.
}
\label{fig:bound}
\end{center}
\end{figure}

We show this constraint in Fig.~\ref{fig:bound} together with predictions of new and
hybrid inflation models to be derived in the following sections. 
We can see that the hybrid inflation model is excluded, while
the new inflation model is on the verge.
It should be noted that, in deriving the constraint (\ref{eq:g-const}),
we have substituted the upper bound on the reheating temperature (\ref{eq:trh-upper}).
Therefore, the constraint on $G_\phi$ becomes severer for lower $T_R$,
proportional to $T_R^{1/2}$.

\section{Single field inflation model}
In this section we estimate $G_{\phi}$ for an inflation model in which the inflaton sector 
consists of a single chiral superfield $\phi$ with nonzero vacuum expectation value (VEV)
in the potential minimum.  In the following 
we assume that the inflaton mass $m_\phi$ is much larger than
the gravitino mass. Since $G_\phi$ corresponds to the fractional
contribution to the SUSY breaking, it is estimated by  minimizing the scalar potential
with the hidden sector responsible for the SUSY breaking.
We assume that the hidden sector contains a chiral superfield $z$ and 
the superpotential is written as
\begin{equation}
   \label{eq:superpot}
   W = W(z) + W(\phi),
\end{equation}
where $W(z)$ and $W(\phi)$ are superpotentials for $z$ and $\phi$,
respectively.
For simplicity we take the minimal K\"ahler potential,
\begin{equation}
   K = |z|^2 + |\phi |^2.
\end{equation}
Then the scalar potential is given by
\begin{equation}
  V = e^G ( |G_z|^2 + |G_\phi|^2 - 3 ).
\end{equation}
Since the cosmological constant should vanish in the true vacuum, 
\begin{equation}
   \label{eq:cosmological-const}
  |G_z|^2 + |G_\phi|^2 = 3 .
\end{equation}
The gravitino mass is given by $m_{3/2} = e^{G/2} \simeq |W|$.
The potential minimum ($=$vacuum) is determined by a condition,
\begin{equation}
  \label{eq:pot_phi}
  V_\phi = e^G ( G_{z\phi} G_{\bar{z}} 
  + G_{\phi\phi}G_{\bar{\phi}} + G_{\phi}) = 0,
\end{equation}
where we have used Eq.~(\ref{eq:cosmological-const}) and
$ G_{i\bar{j}} = \delta_{ij}, (i,j = \phi, z)$.	
Since $z$ is responsible for the SUSY breaking,
$|G_z| \simeq \sqrt{3}$ and  $|W_z| \sim |W| \simeq m_{3/2}$.
On the other hand, $|G_\phi| \ll 1$  for the inflaton field.
Assuming $|G_{\phi}| = |\phi^\dag + W_\phi/W|  \lesssim |\phi|$, we have 
\begin{equation}
   \label{eq:W-W_phi}
   W_{\phi} \sim \phi\, W.
\end{equation}
The order estimation of Eq.~(\ref{eq:pot_phi}) leads to
\begin{equation}
   \phi +\left(\frac{m_\phi}{m_{3/2}}-\phi^2\right) G_{\phi} + G_{\phi} \sim  0,
\end{equation}
where we have omitted the coefficients of order unity, and
used  $W_{\phi\phi} \simeq m_{\phi}$.
Thus we obtain the formula for $G_{\phi}$ in the
single field inflation model,
\begin{equation}
  \label{eq:const-single}
 \la G_{\phi} \ra \sim \frac{m_{3/2}}{m_\phi} \la \phi \ra,
\end{equation}
for $m_\phi \gg m_{3/2}$.
Note that this result satisfies the assumption $|G_{\phi}|  \lesssim |\phi|$, so
our analysis is consistent.
Thus, the inflaton decay into gravitinos can set a 
constraint on a single field inflation model 
when the inflaton $\phi$ takes a non-vanishing 
expectation value after inflation. 

For a concrete example, here we study a new inflation 
model~\cite{Kumekawa:1994gx,Izawa:1996dv,Ibe:2006fs}.
In the new inflation model the superpotential of the inflaton 
sector is written as
\begin{equation}
  W =  v^2\phi - \frac{\phi^{5}}{5},
\end{equation}
where $v = 10^{-7} \sim 10^{-6}$ for producing the observed density
fluctuations. After inflation, the inflaton $\phi$ takes the expectation value
$\sim \sqrt{v}$. In this model the gravitino mass is related to $v$ as
$m_{3/2} \sim v^{5/2}$, and the inflaton mass is given by 
$m_{\phi} \sim v^{3/2}$. Thus, Eq.~(\ref{eq:const-single}) leads to
$G_{\phi} \sim v^{3/2}\sim 6\times 10^{-10}$ for $m_{3/2}=1$~TeV
and $m_{\phi} = 10^9$~GeV, which is close to the constraint shown in
Fig.~\ref{fig:bound}~\footnote{%
The new inflation is also realized for $W = X(v^2 -g \phi^4)$~\cite{Asaka:1999jb}
for which  the constraint can be relaxed. This is because the universe after inflation
is dominated by $\phi$ with suppressed $G_\phi$,
while $G_X$ is not suppressed. This should
be contrasted with the case of the hybrid inflation model.
 }.

\section{Hybrid inflation model}
In the previous section we estimate $G_\phi$ for
 single field inflation models.  However, when
the inflaton sector contains multiple superfields, 
Eq.~(\ref{eq:const-single}) cannot be applied. 
Here, we consider a hybrid inflation model as a representative
example. 

The hybrid inflation model contains two kinds of
superfields: one is $\phi$ which plays a role of inflaton 
and the others are waterfall fields $\Psi$ and 
$\tilde{\Psi}$~\cite{Copeland:1994vg,Dvali:1994ms,Linde:1997sj}. 
After inflation ends, $\phi$ as well as $\Psi$ oscillates around the potential
minimum and dominates the universe until the reheating. 

The total superpotential $W$ is written as
\begin{equation}
   W = W(z) + W(\phi, \Psi,\tilde{\Psi}),
\end{equation}
where the superpotential $W(\phi, \Psi,\tilde{\Psi})$   
for the inflaton sector is
\begin{equation}
   \label{eq:spot_hyb}
   W(\phi, \Psi,\tilde{\Psi}) = \phi (\mu^{2}  
   - \lambda \tilde{\Psi}\Psi).
\end{equation}
Here $\lambda$ is a coupling constant and $\mu$ is the inflation 
energy scale. The potential minimum is located at $\la\phi\ra = 0$
 and $\langle \Psi\rangle = \langle\tilde{\Psi}\rangle =
\mu/\sqrt{\lambda}$ in the SUSY limit. Including the effect of the hidden sector, 
however, the minimum slightly shifts as shown below. 
For successful inflation, $\mu$ and $\lambda$ are related as 
$\mu \simeq 2\times 10^{-3}\lambda^{1/2}$ for
$\lambda \gtrsim 10^{-3}$,
and $\mu \simeq 2\times 10^{-2}\lambda^{5/6}$ for $\lambda \lesssim 10^{-3}$.
Moreover, in this type of hybrid inflation there exists a problem 
of cosmic string formation because $\Psi$ and $\tilde\Psi$ generally have $U(1)$ gauge
charges. To avoid the problem the coupling should be small as, 
$\lambda \sim 10^{-4}$~\cite{Endo:2003fr}.

Now let us estimate $G_{\phi}$ and $G_{\Psi}$. 
The conditions for the potential minimum lead to
\begin{eqnarray}
   G_{z\phi}G_{\bar{z}} + G_{\phi\phi}G_{\bar{\phi}} + G_{\phi}
   + G_{\Psi\phi}G_{\bar{\Psi}} 
   + G_{\tilde{\Psi}\phi}G_{\bar{\tilde{\Psi}}} & = & 0  
   \label{eq:pot-phi-hyb}\\
    G_{z\Psi}G_{\bar{z}} + G_{\phi\Psi}G_{\bar{\phi}} + G_{\Psi}
   + G_{\Psi\Psi}G_{\bar{\Psi}} 
   + G_{\tilde{\Psi}\Psi}G_{\bar{\tilde{\Psi}}} & = & 0.
   \label{eq:pot-psi-hyb}
\end{eqnarray}
Here we do not use the minimization condition for $\tilde{\Psi}$,
since it is equivalent to Eq.~(\ref{eq:pot-psi-hyb}).
In the same way as deriving Eq.~(\ref{eq:W-W_phi}), we assume
$ |G_\Psi | \simeq | G_{\tilde{\Psi}}| \lesssim |\Psi |$, leading to
$W_\Psi/W \sim \Psi$.
Together with  Eq.~(\ref{eq:spot_hyb}), we obtain
\begin{equation}
   \langle \phi \rangle \sim \frac{m_{3/2}}{\lambda},
\end{equation}
where we have used $|W| \simeq m_{3/2}$ and $|\Psi| \simeq |\tilde{\Psi}|$.
Then, with use of  
$G_{\Psi\phi} \sim \lambda\Psi/m_{3/2} - \Psi W_\phi/W$,
$G_{z\Psi} \sim \Psi$, $G_{\Psi\Psi} \sim \Psi^2$
and $G_{\Psi\tilde{\Psi}} \sim \lambda\phi/m_{3/2} - \Psi^2$,
Eqs.~(\ref{eq:pot-phi-hyb}) and (\ref{eq:pot-psi-hyb}) are written as
\begin{eqnarray}
      \frac{W_\phi}{W} + \frac{W_{\phi}^2}{W^2}G_\phi 
      + G_\phi 
      + \left(\frac{\lambda\Psi}{m_{3/2}} 
      - \Psi\frac{W_\phi}{W}\right)G_\Psi
      & \sim & 0  \\
     \Psi 
     + \left(\frac{\lambda\Psi}{m_{3/2}} - \Psi\frac{W_\phi}{W}\right)G_\phi 
      + G_{\Psi} + 
     \Psi^2G_{\Psi}  + 
     \left(\frac{\lambda\phi}{m_{3/2}} - \Psi^2\right)G_\Psi
     & \sim & 0,
\end{eqnarray}
where we omitted coefficients of order unity.
Assuming $|G_\phi| \lesssim |\phi|$, 
we obtain
\begin{eqnarray}
\label{eq:G_phi_hyb}
   \langle G_{\phi}\rangle & \sim & \frac{m_{3/2}}{\lambda}  
   \simeq  \frac{m_{3/2}}{m_{\phi}} \langle \Psi \rangle,
   \\ 
   \langle G_{\Psi}\rangle & \simeq &  \langle G_{\tilde{\Psi}}\rangle \sim
   \frac{m_{3/2}^2}{\lambda^2\langle \Psi \rangle} \simeq 
   \frac{m_{3/2}^2}{m_\phi^2} \langle \Psi \rangle.
   \label{eq:G_psi_hyb}
\end{eqnarray}
Here we have used $W_\phi/W \sim \langle \phi \rangle \sim m_{3/2}/\lambda$
and $m_\phi = \lambda \langle \Psi \rangle$. One can easily check that
the above results satisfy the assumptions we made 
on $G_{\phi}$ and $G_\Psi$. It should be noted that $G_\phi$ is much larger than
$G_\Psi$. Therefore it is $\phi$ that  produces too much gravitinos.

For $m_{3/2}=1$~TeV and $\lambda \sim 1 -10^{-4}$ 
we obtain $\mu \sim 2\times 10^{-3} - 10^{-5}$, 
$G_\phi \sim 4\times 10^{-16}- 4\times 10^{-12}$ 
and $m_{\phi} \sim 5\times 10^{15} - 2 \times 10^{11} $~GeV.
From Fig.~\ref{fig:bound}, one can see the hybrid inflation model is
excluded by the gravitino overproduction. Although the constraint on
$G_\phi$ becomes slightly mild for  $\lambda \lesssim 10^{-4}$,
it is then disfavored by the WMAP result.
This is because the density fluctuation becomes almost scale-invariant  for  $\lambda \lesssim 10^{-4}$
while the spectral index  is $n_s = 0.95 \pm 0.02$  according to the WMAP three year data~\cite{Spergel:2006hy}.

So far we have considered the standard hybrid inflation, but the final results
Eqs.~(\ref{eq:G_phi_hyb}) and (\ref{eq:G_psi_hyb}) also apply to 
a smooth hybrid inflation model~\cite{Lazarides:1995vr}, which is favored compared
to the hybrid inflation model in a sense that 
the predicted spectral index is smaller.
The constraint on this model, however,  is more or less similar to the hybrid inflation,
and the smooth hybrid inflation is also excluded.

\section{Chaotic inflation model}
A chaotic inflation model~\cite{Kawasaki:2000yn,Kawasaki:2000ws} 
is based on a Nambu-Goldstone-like shift symmetry of the
inflaton chiral multiplet $\phi$. Namely, we assume that the
K\"ahler potential $K(\phi,\phi^\dag)$ is invariant under the shift
of $\phi$,
\begin{equation}
  \phi \rightarrow \phi + i\,A,
  \label{eq:shift}
\end{equation}
where $A$ is a dimensionless real parameter. Thus, the K\"ahler
potential is a function of $\phi + \phi^\dag$; $K(\phi,\phi^\dag)
= K(\phi + \phi^\dag)$. We identify its imaginary part 
with the inflaton field $\varphi$.
Moreover, we  introduce a small breaking term of the shift symmetry 
in the superpotential in order for the inflaton $\varphi$ 
to have a potential:
\begin{equation}
  W = mX\phi, 
  \label{eq:mass}
\end{equation}
where we introduced a new chiral multiplet $X$, and $m \simeq 10^{13}$GeV
determines the inflaton mass.
Furthermore, we impose a discrete $Z_2$ symmetry which 
forbids the problematic linear term in K\"{a}hler potential such 
as  $ K= c(\phi + \phi^{\dagger}) + \cdots$\cite{Kawasaki:2000yn,Kawasaki:2000ws}.
Then we take the minimal K\"aher potential,
\begin{equation}
  K =  \frac{1}{2}(\phi + \phi^\dag)^{2} + |X|^2. 
  \label{eq:kahler}
\end{equation}
The superpotential (\ref{eq:mass}) and K\"ahler potential (\ref{eq:kahler})
lead to the scalar potential,
\begin{equation}
    \label{eq:potential4}
     V(\varphi,X) \simeq  \frac{1}{2} m^{2} \varphi^{2} 
            + m^2 |X|^2,
\end{equation}
for $|X| < 1$. For $\varphi \gg 1$ and $|X| < 1$, the $\varphi$ field dominates the
potential and the chaotic inflation takes place [for details see 
Refs~\cite{Kawasaki:2000yn,Kawasaki:2000ws}].

In this chaotic inflation model we see  $G_{\phi} =0$ since $\phi =0$ and $X=0$ 
in the vacuum. Including the hidden sector does not modify the potential minimum due to the $Z_2$ symmetry.
Therefore, the new gravitino problem does not exist.

\section{Conclusions}

In this letter we have shown that an inflation model with a nonzero 
VEV $\la \phi \ra \ne 0$ generically leads to the gravitino overproduction, which
can jeopardize the successful standard cosmology especially when the gravitino 
is unstable and its mass is in the region of $O(100)\,$GeV and $O(10)$\,TeV. 
We have explicitly calculated $G_\phi$, which is an important parameter to determine the gravitino
abundance, for several inflation models.  To be concrete we have fixed  $m_{3/2} = 1\,$TeV and
shown that the  new inflation is on the verge of  being excluded, while the hybrid inflation 
model is excluded. The most attractive way to
get around this new gravitino problem is to have the potential minimum 
at the origin as a class of chaotic inflation models does. Among the known models, such 
a chaotic inflation model 
is favored in a sense that it is free from the potential gravitino 
overproduction problem.

Here, let us comment on other solutions. 
Throughout this letter we have assumed that
no entropy production occurs after the reheating completes. If the huge 
entropy is produced at late time~\cite{Lyth:1995ka}, the new 
gravitino problem can be greatly relaxed.
Another even manifest solution is to assume the gravitino mass 
$m_{3/2} <$ a few keV~\footnote{
Taking account of the constraints from the CMB and the structure formation,
the upper bound on the gravitino mass can be reduced to $O(10)$eV~\cite{Viel:2005qj}. 
}.
In this case, the produced gravitinos get into thermal 
equilibrium due to relatively strong interactions 
with the standard-model particles.

Finally, we shortly note how severe the new gravitino problem is for the gravitino
mass other than $O(1)$TeV.  We will see that it becomes milder for a mass 
region $m_{3/2} < O(1)$ GeV suggested from gauge mediation 
models~\cite{Giudice:1998bp}, since the cosmological constraint is weaker; it 
comes from the requirement that the gravitino abundance should not exceed 
the present dark matter abundance.
For a mass region $m_{3/2} = O(100)$ TeV suggested from  anomaly mediation 
models~\cite{Randall:1998uk}, we will have a  constraint more or less similar 
to the present result. The reason why we have the stringent constraint on $G_\phi$ for
$m_{3/2} = O(100){\rm \,GeV} -O(10)$\,TeV is that the reheating temperature is so severely
constrained by the BBN for this mass range. This means that the reheating temperature 
cannot be arbitrarily small even for  the gravitino
mass other than $O(1)$TeV, since the branching ratio of the direct gravitino production
increases. The extended analysis including a broad mass region for the
gravitino will be given in Ref.~\cite{KTY}. 

\vspace{1cm}
{\bf Note added:}
Very recently, Dine et al~\cite{Dine:2006ii} have pointed out that 
the leading term of $G_\phi$ is cancelled out after taking into 
account a mass mixing between the inflaton and the hidden-sector field $z$
and the effective $G^{\rm eff}_\phi$ becomes much smaller 
if one takes the minimal K\"{a}hler potential. We agree with them in this point. 
However, as pointed out in Ref.~\cite{Dine:2006ii}, 
if the mass of the hidden sector field $z$ is comparable or larger than
the inflaton mass, such a cancellation does not occur. 
In the case of  the new inflation model, the inflaton mass can be comparable to the mass of $z$,
and our constraint in the text is then applicable.
In addition, even if the hidden-sector field mass is smaller than the inflaton mass,
$G_\phi$ may not be suppressed if the K\"{a}hler potential is non-minimal.
In fact, it is quite natural to have a non-minimal K\"{a}hler potential in the 
gravity-mediated SUSY breaking model. If one introduces $\delta K = \kappa/2\, |\phi|^2z^2 + {\rm h.c.}$,
the effective $G^{\rm eff}_\phi$ becomes 
$G^{\rm eff}_\phi \sim \kappa \,m_{3/2}\la\phi\ra/m_\phi $. Thus, $G_\phi$ in the text
should be replaced by this effective $G^{\rm eff}_\phi$. Notice that the hidden-sector
field $z$ has no charge in any symmetry and hence the above K\"{a}hler term has 
no reason to be suppressed.
We find that there is still a very stringent constraint on the hybrid inflation model 
unless the $\kappa$ is extremely small. (For the hybrid inflation model,
$G^{\rm eff}_\phi \sim \kappa \,m_{3/2}\la\Psi\ra/m_\phi$ results from
$\delta K = \kappa/2\,|\Psi|^2z^2 + {\rm h.c.}$, where
$\phi$ and $\Psi$ are the inflaton and  the water fall field, respectively).
The detailed analysis will be given elsewhere.

\section*{Acknowledgments}
F.T. is grateful to Motoi Endo and Koich Hamaguchi
for fruitful discussions. 
F.T.  would like to thank the Japan Society for Promotion of 
Science for financial support. T.T.Y. thanks M. Ibe and Y. Shinbara for a useful discussion.
The work of T.T.Y. has been supported in part by a Humboldt Research Award.


\end{document}